\documentclass[]{pasj02} 
\Received{}
\Accepted{}
\Published{}
\usepackage{url}
\usepackage{comment}
\usepackage{ulem}
\usepackage{amsmath}
\usepackage{booktabs} 
\usepackage{color}
\usepackage{multirow}
\usepackage{graphicx}
\usepackage[switch,mathlines]{lineno}

\begin{document}

\title{X-ray and H$\alpha$ Superflare on an RS CVn-type Star, UX Arietis: Constraint for the Flare Location from Radial Velocity Change during the Flare}

\newcommand{\Add}[1]{\textcolor{red}{\bf #1}}
\newcommand{\Value}[1]{\textcolor{blue}{\bf #1}}
\newcommand{\Ok}[1]{\textcolor{black}{#1}}
\newcommand{\Comment}[1]{\textcolor{orange}{\bf #1}}

\author{
 Sota \textsc{Urabe},\altaffilmark{1}\altemailmark\orcid{} 
 \email{sotaurabe@gmail.com}
 Yohko \textsc{Tsuboi}\altaffilmark{1}\altemailmark\orcid{0000-0001-9943-0024}
 \email{tsuboi@phys.chuo-u.ac.jp}
 Kosuke \textsc{Namekata}\altaffilmark{2,3,4,5}\orcid{0000-0002-1297-9485}
 Sakura \textsc{Nawa}\altaffilmark{1}\orcid{}
 Hiroyuki \textsc{Maehara}\altaffilmark{6}\orcid{0000-0003-0332-0811}
 Noboru \textsc{Nemoto}\altaffilmark{1}\orcid{}
 Yuta \textsc{Notsu}\altaffilmark{7,8}\orcid{0000-0002-0412-0849}
 and 
 Wataru \textsc{Iwakiri}\altaffilmark{1,9}\orcid{0000-0002-0207-9010}
}

\altaffiltext{1}{Department of Physics, Chuo University, 1-13-27 Kasuga, Bunkyo-ku, Tokyo 112-8551, Japan}
\altaffiltext{2}{NASA Goddard Space Flight Center, 8800 Greenbelt Road, Greenbelt, MD 20771, USA}
\altaffiltext{3}{The Catholic University of America, 620 Michigan Avenue, N.E. Washington, DC 20064, USA}
\altaffiltext{4}{The Hakubi Center for Advanced Research/Graduate School of Science, Kyoto University, Kitashirakawa-Oiwake-cho, Sakyo-ku, Kyoto, 606-8502, Japan}
\altaffiltext{5}{Department of Physics, Kyoto University, Kitashirakawa-Oiwake-cho, Sakyo-ku, Kyoto, 606-8502, Japan}
\altaffiltext{6}{Okayama Branch Office, Subaru Telescope, National Astronomical Observatory of Japan, NINS, Kamogata, Asakuchi,
Okayama 719-0232, Japan}
\altaffiltext{7}{Laboratory for Atmospheric and Space Physics, University of Colorado Boulder, 3665 Discovery Drive, Boulder, CO 80303, USA}
\altaffiltext{8}{National Solar Observatory, 3665 Discovery Drive, Boulder, CO 80303, USA}
\altaffiltext{9}{International Center for Hadron Astrophysics, Chiba University, Chiba 263-8522, Japan}


\maketitle

\begin{abstract}

We report on a giant stellar flare from the RS CVn-type binary UX Arietis, detected with the Monitor of All-sky X-ray Image (MAXI) and followed by a 12-day optical spectroscopic campaign using the 3.8~m Seimei Telescope. The flare released $5 \times 10^{37}$~erg in X-rays (0.1--100~keV) and $(2$--$6) \times 10^{36}$~erg in the H$\alpha$ line, placing it among the most energetic events of its kind. The H$\alpha$ light curve showed sinusoidal modulation atop an exponential decay, consistent with reappearance of the flaring region due to binary rotation.

At orbital phase 0, when the primary star is farthest from the observer, 40\% of the H$\alpha$ flux was obscured, while at phase 0.5 the full emission was visible. This suggests the H$\alpha$ emitting region is located at a relatively low latitude and is comparable in size to the stellar disk. Radial velocity modulation implies that the region lies at $\sim19\,R_{\odot}$ from the system's rotation axis, farther out than the stellar limb at $14.4\,R_{\odot}$.

Photometric monitoring with the Chuo-university Astronomical Telescope revealed a large low-latitude starspot covering $\sim25\%$ of the surface. These findings are consistent with a scenario in which the flare occurred above the starspot, and the H$\alpha$-emitting plasma was magnetically confined in a loop extending at least $5\,R_{\odot}$ above the stellar surface.

From the MAXI data and assuming a radiatively cooling plasma, the electron density and volume are estimated to be $10^{10}$~cm$^{-3}$ and $1 \times 10^{35}$~cm$^3$, respectively. If cubic in shape, this corresponds to $7\,R_{\odot}$, consistent with the H$\alpha$ region height.

These results provide direct constraints on the geometry of the plasma and its spatial relationship with the starspot in one of the most energetic stellar flares ever observed.

\end{abstract}


\section{Introduction} \label{sec:intro}

Solar flares are explosive phenomena in which the magnetic energy stored in the solar corona is rapidly released and converted into kinetic, thermal, and non-thermal energies (e.g. \cite{1999ApJ...526L..49S,2011LRSP....8....6S,2002ApJ...577..422S,2015EP&S...67...59M,2017ApJ...851...91N}).
The released energy is transported downward into the lower atmosphere, driving various plasma processes such as chromospheric evaporation.
As a result, hot plasma with temperatures of about $10^7$ K expands upward into the corona along magnetic field loops, forming bright X-ray flare loops.
The largest solar flare ever recorded released an enormous amount of energy, on the order of $10^{32}$ erg (e.g. \cite{2012ApJ...759...71E}).
Stellar flares span a wide range of energies, from small-scale events to some of the most powerful stellar explosions, and are detected across wavelengths from radio to X-rays (e.g., \cite{1974IAUS...57..105K,2017LRSP...14....2B,Haisch+91ARA&A..29..275H}).
By analogy with solar flares, those observed on late-type stars are also thought to be caused by magnetic reconnection occurring in their coronae, as their light curves and spectral evolution are similar to those of solar flares (e.g. \cite{2016PASJ...68...90T,2022PASJ...74..477K}).

With large starspots as well as persistent H$\alpha$ emission, both indicative of enhanced chromospheric activity (e.g. \cite{1990ApJ...348..682S}), RS CVn-type binary systems are known to produce flares that are several orders of magnitude more energetic than solar flares, with released energies ranging from $10^{36}$ to $10^{39}$ erg (e.g. \cite{2016PASJ...68...90T,2022PASJ...74..477K,2023ApJ...948....9I,2025A&A...695L...2C}).

The large radiative energies may be released via an extensive magnetic structures, for example magnetic loops connecting the binary components (e.g. \cite{1980ApJ...239..911S,1985IAUS..107..281U}).
To understand the geometry of flare loops in RS CVn binaries, estimates of loop lengths have been made using X-ray studies, comparing them with stellar radius and binary separations (e.g. \cite{2016PASJ...68...90T,2021ApJ...910...25S,2022PASJ...74..477K}). These estimates have been obtained through spectral and timing analyzes using methods such as those described by \citet{1999ApJ...526L..49S,2002ApJ...577..422S}, \citet{1983ASSL..102..255H}, \citet{1989A&A...213..245V}, \citet{1997AA...325..782R}, and \citet{2007AA...471..271R}. For example, \citet{2016PASJ...68...90T} analyzed 18 MAXI X-ray flares from eight RS CVn systems and estimated loop lengths ranging from 0.4 to 20 times the binary separation, using the scaling relations of \citet{1999ApJ...526L..49S,2002ApJ...577..422S} and a pre-flare coronal density of $n_{\rm e} = 10^9$ cm$^{-3}$. \citet{2022PASJ...74..477K} also estimated loop lengths of 3.1 to 3.3 binary separations using \citet{1999ApJ...526L..49S,2002ApJ...577..422S}'s scaling relation, and 0.6 to 1.3 binary separations using \citet{1983ASSL..102..255H}'s scaling relation. However, despite the suggestions of the large scale of flaring loops, it is not conclusive whether the loops are connected with each binary components, since there should be systematic uncertainties in the physical parameters such as preflare electron densities or the shape of the loops.

In this case, (1) high-dispersion spectroscopic observations and/or (2) multi-wavelengths observations in giant flares may be useful; the former can capture the dynamism of the emission region and the latter can resolve the different temperature/spatial regions of the same event in the different bands, thus providing information comparable to spatial information in stellar flares that are distant from us and lack spatial resolution. However, giant flares in stars occur infrequently, and although multi-wavelength observations exist (e.g. \cite{2022PASJ...74..477K}), they are still scarce.

Here, we present follow-up mid-resolution H$\alpha$ spectroscopic observations conducted using the Japanese Seimei telescope at Okayama Observatory, triggered by MAXI's detection of a large stellar superflare from the RS CVn-type star UX Ari. The observations were carried out over 12 days, beginning on April 3, 2022. The superflare lasted several days and exhibited rotational radial velocity modulation in the H$\alpha$ line during the follow-up observations, which is the primary focus of this study. Additionally, we conducted long-term photometric observations of UX Ari using the Chuo-university Astronomical Telescope (CAT). The rotational modulations of stellar brightness and H$\alpha$ RV changes allow us to estimate the relationship between starspots/active regions and the giant flare. In this paper, we provide an overview of the observations in Section \ref{sec:obs}, present the analysis and results in Section \ref{sec:res}, discuss the flare radiation energetics, loop length scale, and potential flare locations in Section \ref{sec:discussion}, and conclude with a summary in Section \ref{sec:summary}.

\section{Observation} \label{sec:obs}
\subsection{Target Star: UX Ari} \label{sec:obs:target}
UX Arietis (UX Ari, HD 21242, HIP 16042) is classified as an RS CVn-type binary system which is synchronized in their spin period and orbital period.
It is located at a distance of 50.6 pc from Earth, based on the Gaia EDR3 parallax (\cite{GaiaEDR3_2021,BailerJones2021}).
The binary properties are summarized in Table \ref{tab:parameter}.
It consists of the primary K0-type subgiant (radius = 5.6 ± 0.1 {$R_{\odot}$}) and the secondary G5-type main-sequence star (radius = 1.6 ± 0.2 {$R_{\odot}$})
\citep{2017ApJ...844..115H}.
The distance between the primary and secondary stars is as close as 18.8 ± 0.1 {$R_{\odot}$}, and the inclination angle is 125.0 ± 0.5 degrees. The orbital/spin period is reported to be 6.437888 days \citep{2017ApJ...844..115H}.
Among RS CVn-type binaries, UX Ari is especially well-known as a very active flare star, with many reports of giant flares. \citep{1978BAAS...10..418F, 1989PASJ...41..679T, 1998A&A...332..149M, 2002evn..conf..275M, 2003csss...12..981C, 2005A&A...435L...1M, 2011PhDT........43P, 2016PASJ...68...90T, 2022PASJ...74..477K}.
For example, \citep{2025A&A...695L...2C} reported that a Doppler shift appeared as an absorption component in the H$\alpha$ line during the flare, which was interpreted as a filament eruption.

\begin{table}[htbp]
  \caption{Stellar properties of the UX Ari system.}
  \label{tab:parameter}
  \centering
  \begin{tabular}{cccc}
    \hline
    Parameters & Primary & Secondary & Ref. \\
    \hline
    Spectral type & K0IV & G5V & (1) \\
    Radius [$R_{\odot}$] & 5.6$\pm$0.1 & 1.6$\pm$0.2 & (2) \\
    Mass [$M_{\odot}$] & 1.30$\pm$0.06 & 1.14$\pm$0.06 & (2) \\
    Effective temperature [K] & 4560$\pm$100 & 5670$\pm$100 & (2) \\
    Spot temperature [K] & 3120--4100 & -- & (2) \\
    Separation [$R_{\odot}$] & \multicolumn{2}{c}{18.8$\pm$0.1} & (2) \\
    Distance [pc] & \multicolumn{2}{c}{50.6} & (3)(4) \\
    Inclination [degree] & \multicolumn{2}{c}{125.0$\pm$0.5} & (2) \\
    Orbital period [day] & \multicolumn{2}{c}{6.437888$\pm$0.000007} & (2) \\
    \hline
  \end{tabular}
  \vspace{2mm}
  \begin{flushleft}
    \textbf{Note. } References: (1) \citet{1971PASP...83..504C}, (2) \citet{2017ApJ...844..115H}, (3) \citet{GaiaEDR3_2021}, (4) \citet{BailerJones2021}
  \end{flushleft}
\end{table}

\subsection{MAXI} \label{sec:obs:maxi}
On 2022 April 3 at 04:33:24 UTC, an X-ray flare from UX Ari was detected with MAXI\footnote{\url{http://maxi.riken.jp/alert/novae/9672184033/9672184033.htm}} (\cite{2009PASJ...61..999M}).
In this work, we use data from MJD 59670.0 (UTC 0:00 on 2022 April 1)–59687.8 (UTC 19:12 on 2022 April 1) to cover the long duration of this flare.
MAXI is an astronomical X-ray observatory onboard the International Space Station (ISS). In this work, we used data from the Gas Slit Camera, which is sensitive in the 2–30 keV band.
The field of view of these cameras is 160$^{\circ}$×3$^{\circ}$, corresponding to approximately 2\% of the whole sky, scanning roughly 85\% of the entire sky per ISS orbit.

The time interval of MAXI observations is about 92 minutes, which coincides with the orbital period of the ISS.
The exposure time in each ISS orbit was 5–50 seconds in this observation.
We used the “nova alerts” system (\cite{2016PASJ...68S...1N}) to perform the follow-up observations of MAXI-detected stellar flares (see, e.g. \cite{2022PASJ...74..477K} for the detailed explanations of the follow-up system).
The alert system has been previously used to search large flares from UX Ari (e.g. \cite{2016PASJ...68...90T,2022PASJ...74..477K}).

\subsection{Seimei Telescope} \label{sec:obs:seimei}
Optical spectroscopic observations of UX Ari were carried out using the Kyoto Okayama Optical Low-dispersion Spectrograph with Optical Fiber Integral Field Unit (KOOLS-IFU; \cite{2019PASJ...71..102M}), mounted on the 3.8 m Seimei telescope (\cite{2020PASJ...72...48K}). This instrument provides a mid-resolution optical spectra with a wavelength coverage from 5800 to 8000 {\AA}, including the hydrogen H$\alpha$ line (6562.8 {\AA}), and a wavelength resolution R ($\lambda$/$\Delta\lambda$) of approximately 2000.
The first observation by the Seimei telescope started 5 hours and 29 minutes after the MAXI detection of a gigantic X-ray flare at 4:33:24 UT on 2022 April 3. Following the initial observation, 19-days follow-up observations were conducted to characterize the quiescent component and rotation variations.
The exposure time for each frame was set to be 30 s for these nights. We took 3–49 frames per day.

\begin{figure}[htbp]
\begin{center}
  \includegraphics[width=0.9\linewidth]{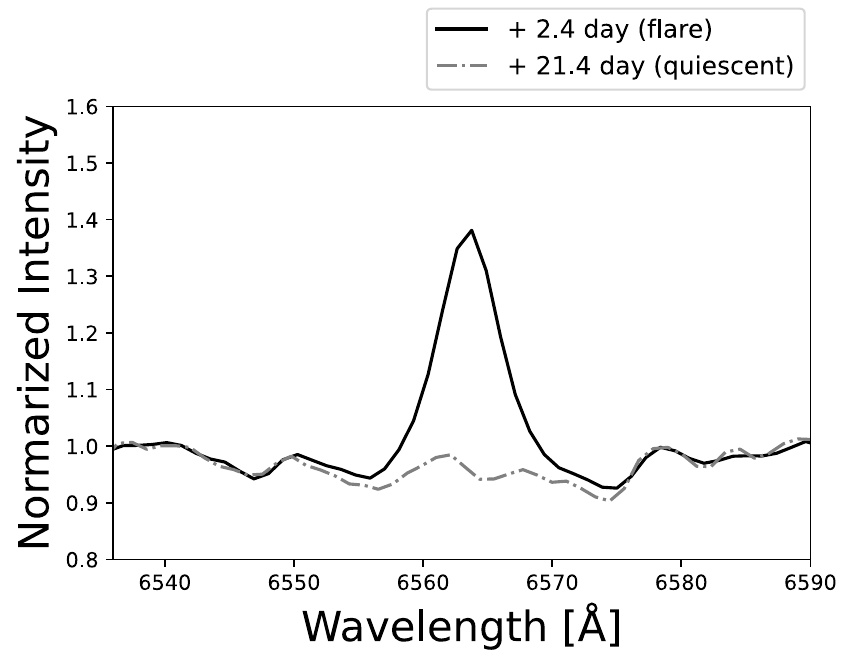}
\end{center}
  \caption{One example of the H$\alpha$ spectra normalized by continuum level. A black solid line represents a flaring spectrum on 2022 April 3, while a gray line is a quiescent spectrum obtained in the post flare phase of 2022 April 22. The standard date in the legend is from MJD 59670 as in Figure~\ref{fig:lc}.\\
  {Alt text: Line graph comparing two H$\alpha$ spectra during flare and quiescent phases.}}
  \label{fig:1}
\end{figure}

Data reduction was conducted in accordance with the methodology outlined in \citet{2020PASJ...72...68N}, \citet{2022NatAs...6..241N}, and \citet{2022ApJ...926L...5N}, utilizing both the \textsf{IRAF}\footnote{\textsf{IRAF} is distributed by the National Optical Astronomy Observatories, which are operated by the Association of Universities for Research in Astronomy, Inc., under cooperative agreement with the National Science Foundation.} and \textsf{PyRAF}\footnote{\textsf{PyRAF} is part of the stscipython package of astronomical data analysis tools and is a product of the Science Software Branch at the Space Telescope Science Institute.} software packages.
This process includes bias subtraction, cosmic-ray elimination, sky background subtraction, and wavelength correction for Earth's motion, specifically via a heliocentric correction using the \textsf{rvcorrect} function.
It should be noted that the proper motion of UX Ari was not accounted for in this correction process.
While the sky background was subtracted by employing the spectra of the sky fibers, any data presenting high levels of background light and significant noise were excluded from our analysis.
The broadband continuum spectra were fitted and normalized by using the \textsf{continuum} function in \textsf{IRAF}.
After this normalization, we measure the H$\alpha$ equivalent width (“EW”, H$\alpha$ emission integrated for 6562.8--20 {\AA} $\sim$ 6562.8+20 {\AA}). Figure \ref{fig:1} is an example of the H$\alpha$ spectrum of UX Ari during and after the reported flare.

\subsection{CAT} \label{sec:obs:cat}
The optical photometric observation were performed using the Chuo-university Astronomical Telescope (CAT). 
CAT is located on the rooftop of Chuo University's campus in Tokyo (latitude = 35$\arcdeg$42$\arcmin$30$\arcsec$ N, longitude = 139$\arcdeg$44$\arcmin$54$\arcsec$ E). 

CAT is a 26 cm diameter Vixen Original Maksutov Cassegrain telescope. 
The imaging camera is the SBIG STL-6303E CCD camera, with a field of view of approximately 0.9° × 0.6° with a pixel number of 3072 × 2048 pixels. Four different filters are installed, including V-band (central wavelength: 5380 {\AA}, bandwidth: 980 {\AA}), R-band (central wavelength: 6300 {\AA}, bandwidth: 1180 {\AA}), I-band (central wavelength: 8940 {\AA}, bandwidth: 3370 {\AA}), and H$\alpha$-band (central wavelength: 6563 {\AA}, bandwidth: 1 {\AA})\citep{2005ARA&A..43..293B}.

In this paper, we report the observing results of the V-band photometric data of UX Ari, which were obtained between 2022 January 1 and 2022 March 15, totaling 28 nights.  
The exposure time allocations used in the data were as follows: 0.5 sec (17 nights), 1 sec (8 nights), 1.5 sec (2 nights), and 2.5 sec (1 night). 
Basic correction including dark and flat-field corrections were applied in the analysis. 
Aperture photometry was employed, and the analysis was performed using \textsf{photutils} (version 1.4.0) in python. 
HD 21062 (A-type star, V-band magnitude of 7.12; \cite{2000A&A...355L..27H}) was used as the standard star for absolute photometric calibration. The flux of UX Ari was converted into the apparent V-band magnitude by comparing it with the flux of HD 21062, whose magnitude is well established.

\section{Results} \label{sec:res}
\subsection{Flare occurred at MJD 59672} \label{sec:res:lc}
\begin{figure*}[t]
\begin{center}
  \includegraphics[width=\linewidth]{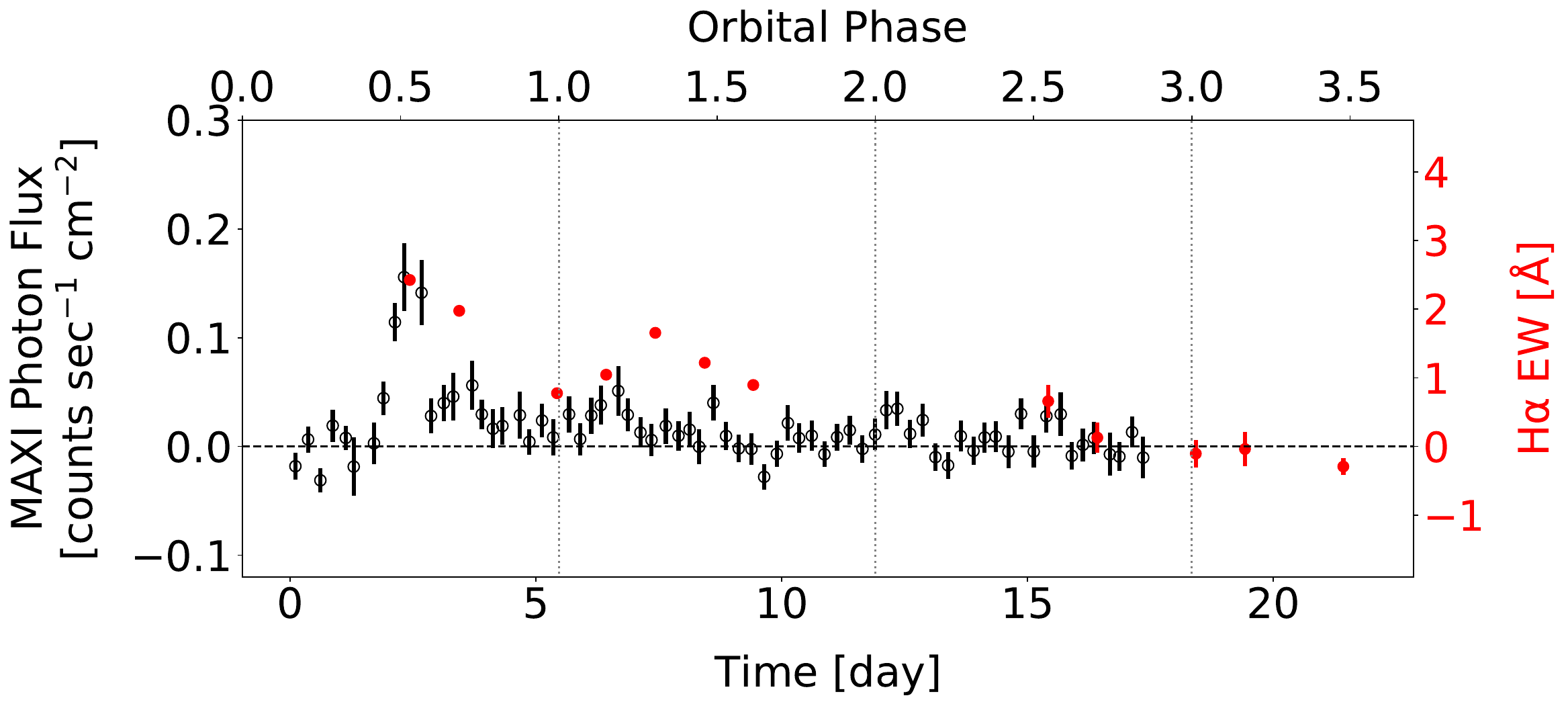}
\end{center}
  \caption{Light curve of a superflare in the X-ray and H$\alpha$ line bands. Black open circles with error bars represent the count rate in the 2.0-6.0 keV band observed with MAXI, while red filled circles with error bars represent the equivalent width of H$\alpha$ line observed with Seimei telescope. The error bars for the first seven data points for Seimei are too small and are buried within the symbol. All error bars represent 1 sigma confidence level. The bottom horizontal axis shows time from April 1, 2022 0:00 UT (= MJD-59670.0). The top horizontal axis indicates the phase of the orbital period (Orbital Period = 6.437888 days).\\
  Alt text: Line graph showing time variation of X-ray and H$\alpha$ emission}
  \label{fig:lc}
\end{figure*}

Figure \ref{fig:lc} shows the observed X-ray light curve.  
The flare peak appears around MJD 59672.5, although the data is sparse.
Approximately 5.5 hours later, at 10:02:25 UT on 2022 April 3 ( MJD 59672.41835), follow-up observations started using the 3.8 m Seimei telescope. 
Subsequent monitoring with the Seimei telescope continued for a period considerably longer than the rotation and orbital period of 6.4 days \citep{2017ApJ...844..115H}, monitoring the object until it nearly returned to its quiescent phase (i.e., H$\alpha$ EW $\sim$ 0 {\AA}).
A clear secondary peak can be seen in the H$\alpha$ light curve around 7.5 days in Figure \ref{fig:lc}.
On the other hand, there is no significant increase in X-rays around 7.5 days, although there is a possible tiny bump-like feature. 

\subsection{X-ray Analysis} \label{sec:res:xray}
\begin{figure}[htbp]
\begin{center}
  \includegraphics[width=0.9\linewidth]{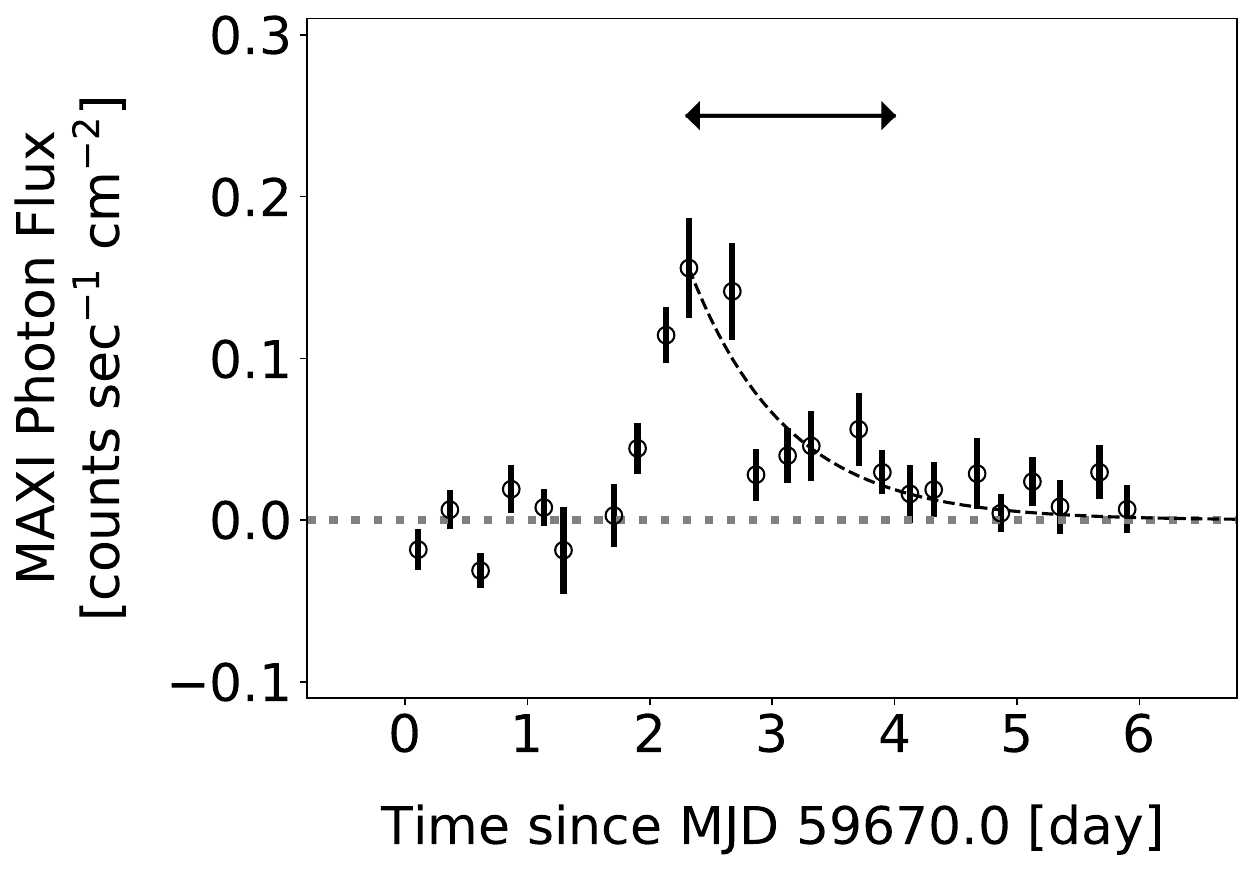}
\end{center}
  \caption{Enlarged view of the light curve of the superflare in the 2.0--6.0 keV band obtained with MAXI. The data points are binned by 4 orbits. The black dashed line is the best fit model for the exponential decay model. The black arrow represents the range used for the fitting.\\
  Alt text: Line graph of an X-ray light curve with an exponential decay fit.}
  \label{fig:lcfit_x}
\end{figure}

To obtain X-ray decay time, we performed a fitting with an exponential model using \textsf{lcurve}\footnote{\textsf{lcurve} is included in HEASoft's ftools package developed and maintained by NASA. It can draw light curves.}. 
Such an exponential decay has been commonly observed in the cooling phases of both solar and stellar flares 
(e.g. \cite{1997A&A...325..782R,2016PASJ...68...90T}).
Since the peak of the flare is seen at MJD 59672.5, we used the data during MJD 59672.3--59674.0 days for the fitting. In the fitting, we fixed the MAXI photon flux at the peak to 0.156 counts sec$^{-1}$ cm$^{-2}$, which is the average value at the peak.
As a result, we obtained the $e$-folding decay time of $70\pm 30$ ksec. The best-fit model is shown in Figure \ref{fig:lcfit_x}.  

\begin{figure}[htbp]
\begin{center}
  \includegraphics[width=1.0\linewidth]{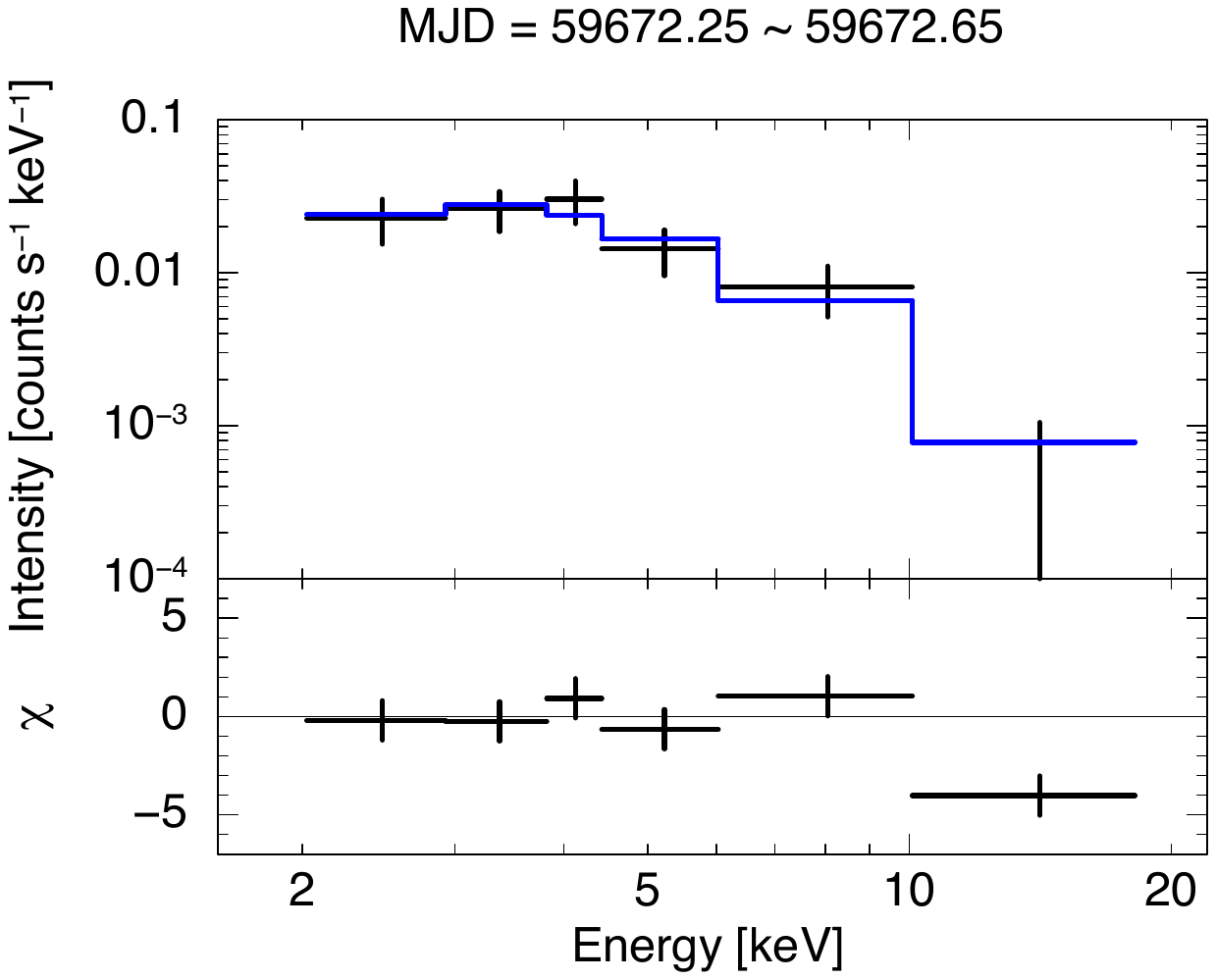}
\end{center}
  \caption{X-ray spectrum observed with MAXI during the flare peak of MJD 59672.25–59672.65.
(Upper) The data points with error bars are the observed X-ray spectrum and the blue line is the best-fit model spectrum. (Lower) Residuals for the best-fit model. Error bards represent 90 \% confidence level.\\
Alt text: Two-panel graph showing an X-ray spectrum with a model fit and residuals.}
  \label{fig:spec_x}
\end{figure}

Figure \ref{fig:spec_x} shows the background-subtracted X-ray spectrum of UX Ari obtained with MAXI during MJD 59672.25–59672.65.
We fitted the data with a single-temperature coronal model, \textsf{apec}, in \textsf{xspec}\footnote{\textsf{xspec} is a package for spectral analysis included in HEASoft, which is developed and managed by NASA.}.
The \textsf{apec} model is one of the optically thin, collisionally ionized plasma model in thermal equilibrium (\cite{2001ApJ...556L..91S}). Optically thin thermal plasma model is composed of thermal bremsstrahlung and line emission. In the apec model, the atomDB is used (\cite{2012ApJ...756..128F}).
In the fitting, we fixed the redshift at 0.0 km s$^{-1}$ and metal abundance at 0.3 times the solar abundance (e.g. \cite{Imanishi+01ApJ...557..747I}), and used \textsf{C-statics} for statistics, since the photon statistics are small.

With the C-statistics of 1.78, in degree of the freedom of 4, we obtained the best-fit parameters.
We show the best-fit model as the blue solid line in Figure \ref{fig:spec_x}, and 
summarize the best-fit parameters in Table \ref{tab:2}.
Multiplying the peak X-ray luminosity and $e$-folding time, we estimated the flare energy as $3 \pm 2 \times 10^{37}$ erg, as indicated in Table \ref{tab:2}.

\begin{table*}[t]
\caption{Physical parameters obtained with light curve and spectral analyses.}
\label{tab:2}
\centering
\begin{tabular}{ccccccccc}
\hline
$t_{\mathrm{H}\alpha}$ & $E_{\mathrm{H}\alpha}$ & $t_{\mathrm{H}\alpha}$ & $E_{\mathrm{H}\alpha}$ & $t_X$ & $EM$ & $T$ & $E_X$ & $E_{X,\mathrm{bol}}$ \\
\multicolumn{2}{c}{(exponential)} & \multicolumn{2}{c}{(exp$\times$sin; Eq.~\ref{eq:fit})} & & & & (2.0--6.0 keV) & (0.1--100 keV) \\
{[ks]} & [$10^{36}$erg] & {[ks]} & [$10^{36}$erg] & {[ks]} & [$10^{55}$ cm$^{-3}$] & [$10^{7}$K] & [$10^{37}$erg] & [$10^{37}$erg] \\
\hline
200 & 2 & 600 & 6 & 70 & 5 & 5 & 3 & 5 \\
(100--400) & (1--3) & (400--800) & (4--8) & (40--100) & (3--6) & (3--8) & (1--5) & (3--9) \\
\hline
\end{tabular}
\vspace{2mm}
\begin{flushleft}
\textbf{Note. } $E_{\rm H\alpha}$, $E_{\rm X}$, $E_{\rm X,bol}$ are radiative energy in the H$\alpha$ line, in the 2.0--6.0 keV band, and in the 0.1--100 keV band, respectively. $t_{\rm H\alpha}$ and $t_{\rm X}$
are {\it e}-folding decay time in the H$\alpha$ line and in the 2.0--6.0 keV band, respectively. 
The physical properties obtained from the H$\alpha$ line are shown separately in the two decaying models. $EM$ and $T$ are the best-fit emission measure and plasma temperature, respectively, obtained with MAXI data during the flare peak of MJD 59672.25–59672.65.
All the errors indicated here are 90 \% confidence range.
\end{flushleft}
\end{table*}

\subsection{H$\alpha$ Analysis} \label{sec:res:ha}
\begin{figure}[htbp]
\begin{center}
\includegraphics[width=1.0\linewidth]{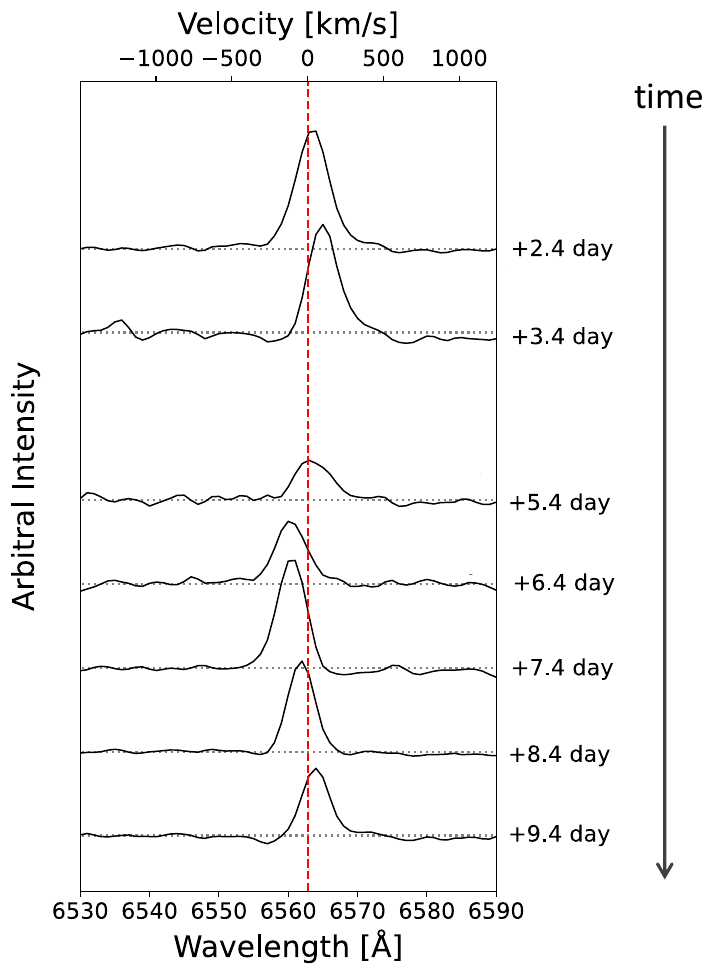}
\end{center}
  \caption{Daily variations of the H$\alpha$ spectra from April 3 (top) to 2022 April 10 (bottom). The dashed line represents the central wavelength of the H$\alpha$ line in the stationary state, at 6562.8 {\AA}. Dates indicate the time elapsed since MJD 59670.0.\\
  Alt text: Line graphs showing daily H$\alpha$ spectral variations}
  \label{fig:spec_ha}
\end{figure}

Figure \ref{fig:spec_ha} shows the spectra near the H$\alpha$ line in daily basis during the period from MJD 59672.4 to MJD 59679.4.
We can see the daily changes both in the intensity and in the central wavelength.
By fitting the spectra with a single Gaussian function, we determined the EW and the central wavelength, respectively.
We report the results from the EW values in this section, while we discuss the central wavelength values in Section \ref{sec:dis:loc}.

\begin{figure*}[t]
\begin{center}
\includegraphics[width=1.0\linewidth]{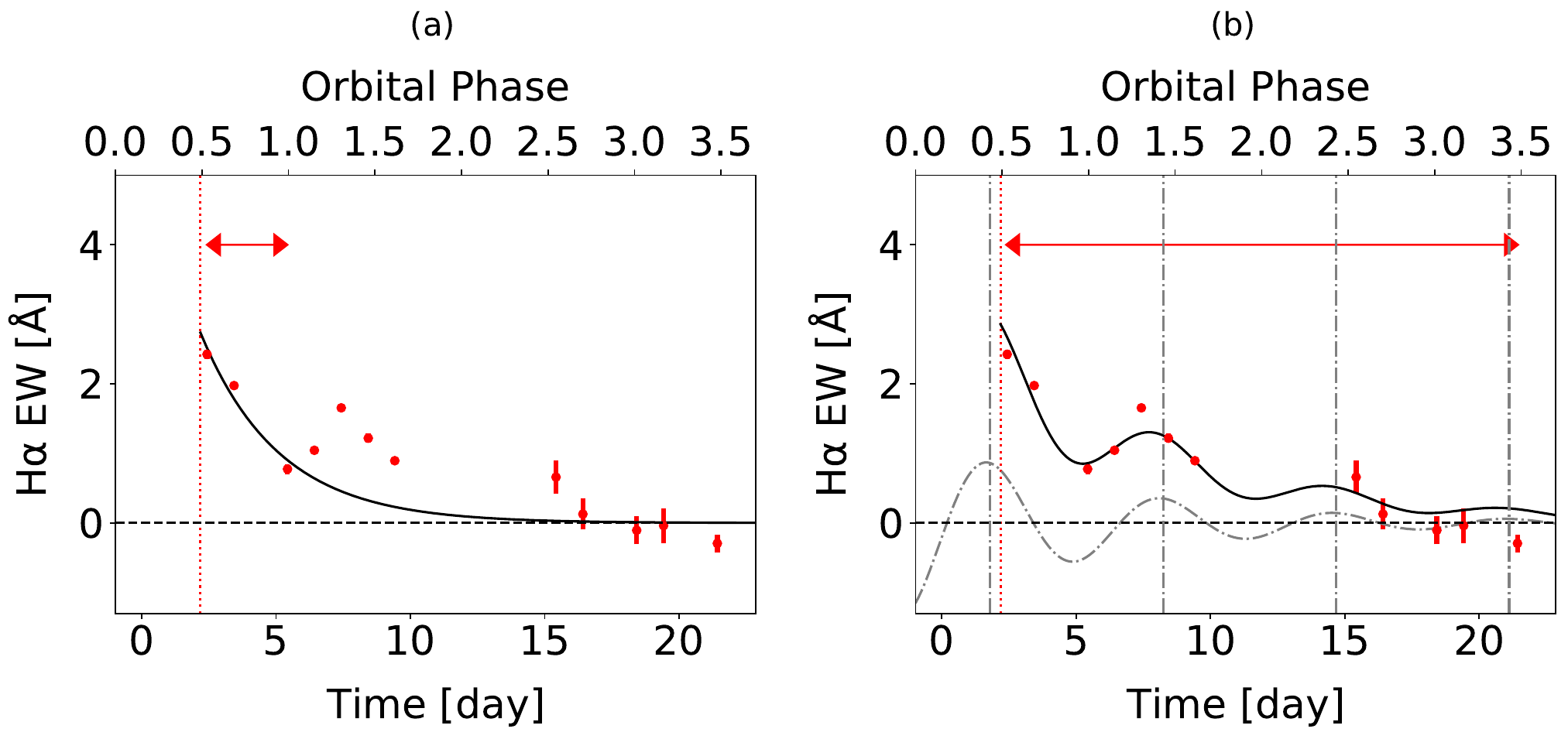}
\end{center}
  \caption{Two fitting results done for the light curve in H$\alpha$ line band (Figure \ref{fig:lc}). Solid lines indicate the best fit models. Horizontal arrows indicate the respective fitting range. Vertical dashed line indicates the assumed time when the flare occurred (i.e., MAXI's X-ray flare trigger time 2022 April 3 at 04:33:24 UTC). (a) A single exponential function (b) A combination of the exponential and sinusoidal functions (Equation \ref{eq:fit}). The gray chained line in panel (b) shows the sinusoidal component (the second term in Equation 1) in the best fit model and its inflection point.\\
  Alt text: Two-panel graph showing H$\alpha$ light curve with model fits using exponential and sinusoidal functions.}
  \label{fig:lcfit_ha}
\end{figure*}

As noted in Section \ref{sec:res:lc}, there is the secondary bump in the H$\alpha$ light curve. 
This can be attributable to the following scenarios: (a) a second flare occurrence or (b) the obscuration and revelation of a single flare due to stellar rotation. 
For both scenarios, we used the following two different fitting approaches, respectively: (I) fitting only the decay of the initial peak in 2.4–5.4 days from MJD 59670.0 with a single exponential function, and (I\hspace{-1.2pt}I) fitting all data (from MJD 59672.4 to MJD 59691.4) with a combined function of exponential and sinusoidal forms. 
For the fitting (I\hspace{-1.2pt}I), we employed the function expressed as,
\begin{eqnarray}\label{eq:fit}
A \exp \left(- \frac{t}{\tau}\right) \left(1+B \sin\left( \frac{2\pi t}{P_{\rm rot}} + \phi\right)\right),
\end{eqnarray}
where $A$ is the normalization constant, $B$ is the amplitude of the sinusoidal modulation, $\tau$ is the {\it e}-folding decay time, $\phi$ is the initial phase, and $P_{\rm rot}$ is the orbital period of the system. 
Figure \ref{fig:lcfit_ha} shows both fitting results. 
In the case of single exponential decay, we obtained a best-fit {\it e}-folding decay time of $300\pm100$ ksec, while in the case of exponential-sinusoidal decay, we obtained a {\it e}-folding decay time of $600\pm200$ ksec.

Assuming that the peak time of H$\alpha$ coincides with that of the X-ray, we estimated each of the peak H$\alpha$ EW of \Ok{2.7} {\AA} and \Ok{2.9} {\AA} for fitting (I) and (I\hspace{-1.2pt}I), respectively.
Deriving the peak H$\alpha$ intensity from the peak EW and the continuum flux, and
multiplying it and {\it e}-folding decay time, we derived the radiative energy of the H$\alpha$ flare emission as $2\pm 1 \times$10$^{36}$ erg and $6\pm 2 \times$10$^{36}$ erg for the fitting (I) and (I\hspace{-1.2pt}I), respectively. 
We note that the continuum flux was calculated based on the R-band magnitude of UX Ari (5.74 magnutude; \cite{2004Ap.....47..443A}) and R-band flux of Vega ($1.8\times10^{-9}$ erg s$^{-1}$ cm$^{-2}$ {\AA}$^{-1}$)\footnote{\url{https://www.stsci.edu/instruments/observatory/PDF/scs8.rev.pdf}}. We summarize these $e$-folding decay time and energy in Table \ref{tab:2}.

\subsection{V-band Variation in Quiescent Phase}\label{sec:res:rotation}
\begin{figure}[htbp]
\begin{center}
\includegraphics[width=1.0\linewidth]{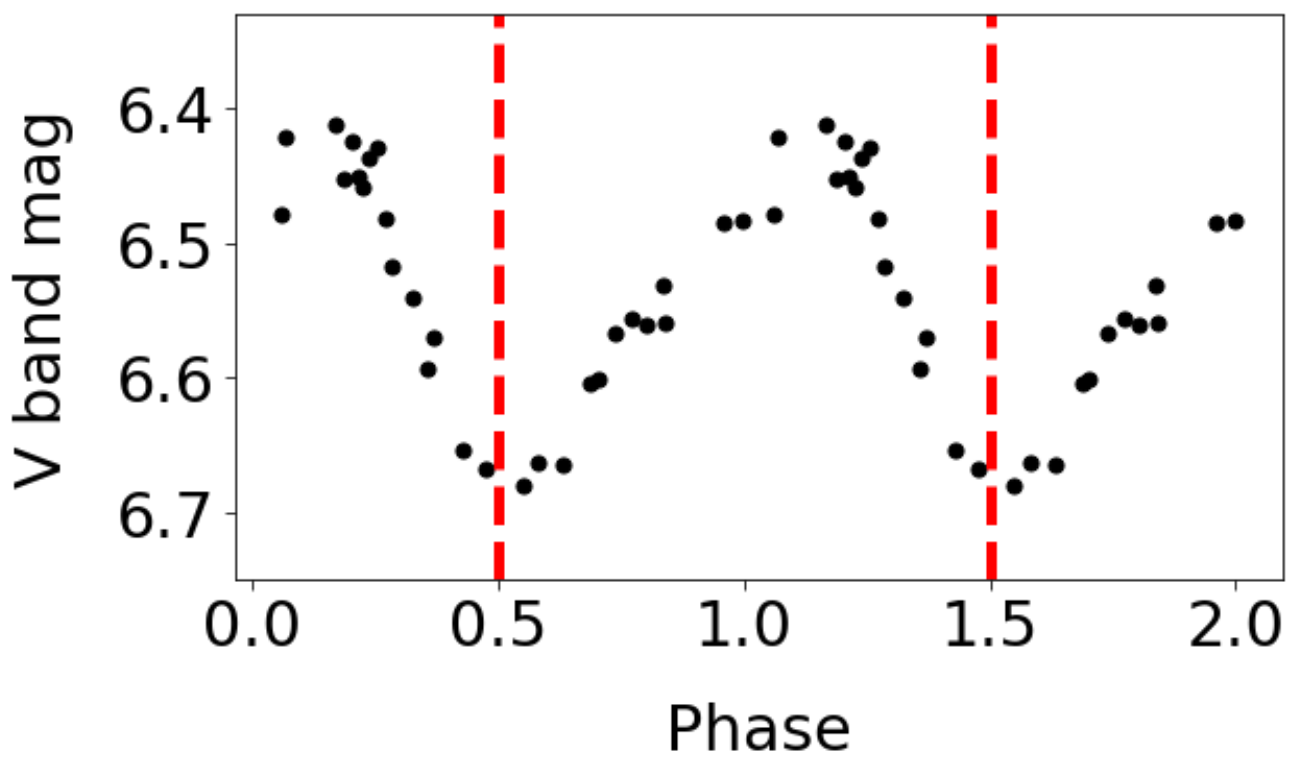}
\end{center}
  \caption{Phase-folded light curve of UX Ari in the V-band magnitude. In the figure, the same folded curves are shown repeatedly. Orbital period is assumed to be 6.437888 days. The V-band magnitude is a relative magnitude from the standard star, HD 21062. The data were obtained with CAT for the period between January 1 to March 15, 2022. Data are averaged by one night. Vertical dashed lines indicate phase of 0.5 and 1.5.\\
  Alt text: Line graph showing phase-folded V-band light curve of UX Ari}
  \label{fig:rot-CAT}
\end{figure}

Figure \ref{fig:rot-CAT} shows the phase-folded V-band magnitude taken with CAT during the period between 2022 January 1 and 2022 March 15. Here, the V-band magnitudes were averaged on daily basis, and an orbital/spin period is supposed as 6.437888 \citep{2017ApJ...844..115H}.
As seen in \citet{2017ApJ...844..115H}, the ephemeris is defined as follows; when the primary star is directly in front of the observer, the rotation phase is 0.5. The each phase exhibits the error of just 0.0006 days, i.e. almost negligible to the orbital period of 6.437888 days. The error was obtained by summing the error of the orbital period of 0.000007 days \citep{2017ApJ...844..115H} by the number of the rotation which was done between the observation in 2012 by \citet{2017ApJ...844..115H} and our observation. 

As can be seen, there is a clear single-peaked modulation in Figure \ref{fig:rot-CAT}.
This can be interpreted as a large starspot or a group of starspots on the surface of one of the components, repeatedly appearing and disappearing as the star rotates (e.g. \cite{2017ApJ...844..115H}).
Hereafter, we call the area ``the spot area''. Here, we are using the phase definition by \citet{2017ApJ...844..115H}. Doppler imaging based on high-resolution spectroscopy has revealed stellar spots on the surface of the subgiant primary of UX Ari (\cite{1991LNP...380..297V, 1999anot.conf..222A, 2004IAUS..219..873G, 2017ApJ...844..115H}). On the other hand, in Figure \ref{fig:rot-CAT}, the flux maximum is near phase 0, while the minimum is near phase 0.5. These indicate that ``the spot area'' is on the primary star side and on the farther location from the secondary star.

Next, we estimate the occupancy of “the spot area”. \citet{2014Ap.....57..344A} monitored the V-band brightness over the period from 1970 to 2010,
as in Figure 4 of their paper. In the light curve, long-term variations can be seen in addition to the short-term rotational modulation. 
During the active phase observed from around 1985 to 2000, particularly near 1989 and 1994, the minimum V-band magnitude of 6.35 was recorded at the brightest phase of the short-term rotational modulation.
We assume that at the flux-maximum, there are almost no star spots on the hemisphere of the primary. On the other hand, our derived least V-band magnitude is about 6.40, which is almost the same as the least V-band magnitude during the long monitoring by \citet{2014Ap.....57..344A}.
This means that even in our observation, there are almost no star spots on the hemisphere of the primary star during the flux-maximum phase.


From now, we derive “the spot area” occupancy using our obtained largest magnitude of 6.68.
We assume that the difference between our magnitude and the least V-band magnitude of 6.35 in \citet{2014Ap.....57..344A} comes from the appearance of the spot area'', and then we estimate the ratio of the spot area'' ($A_{\text{spot}}$) and the hemispheric area of the primary star ($A_{\text{star}}$).
Using the difference between the flux maximum ($F_{\max}$) and the flux minimum ($F_{\min}$), we can express the following equation;

\begin{equation} \label{eq:spot3}
\begin{aligned}
\frac{\Delta F}{F_{\text{max}}} 
&= \frac{F_{\text{max}} - F_{\text{min}}}{F_{\text{max}}} \\
&= \frac{A_{\text{spot}} \left( 
\displaystyle\int_{\text{V-band}} B(\lambda, T_{\text{star}})\, d\lambda 
- \int_{\text{V-band}} B(\lambda, T_{\text{spot}})\, d\lambda \right)}
{A_{\text{star}} \displaystyle\int_{\text{V-band}} B(\lambda, T_{\text{star}})\, d\lambda}
\end{aligned}
\end{equation}

Here, $B\left(\lambda, T_{\text {star }}\right)$ is the radiance of the Black body and is given by

\begin{equation}\label{eq:spot4}
B(\lambda, T)=\frac{2 h c^2}{\lambda^5} \cdot \frac{1}{\exp (h c / \lambda k T)-1} 
\end{equation}

, and $T_{\text{star}}$ and $T_{\text{spot}}$ represent the photosphere temperature and the spot temperature, respectively.
From Equation \ref{eq:spot3}, we obtained the value of “the spot area” occupancy ($A_{\text{spot}}$ / $A_{\text{star}}$) as 25\% of the hemispherical surface at the largest magnitude phase during our observation.
Here, we used the same assumption as \citet{2014Ap.....57..344A}, i.e. we used 4750 K for $T_{\text{star}}$ from \citet{1991LNP...380..297V} and 3450 K for $T_{\text{spot}}$, assuming that $T_{\text{spot}}$ is cooler than $T_{\text{star}}$ by 1300 K (\cite{2014Ap.....57..344A}).
The range for the integration is 4890--5870 {\AA} (see also Section \ref{sec:obs:cat}).

Furthermore, we have confirmed the validity of our method in determining “the spot area” occupancy, using long monitoring by \citet{2014Ap.....57..344A}. During the period around 1982--1986, the maximum and the minimum V-band magnitude are almost the same as ours, as seen in their Figure 4, and the authors estimated the spot occupancy as 20--50 \% of the hemispheric area (10--25 \% of the total area), which is not inconsistent with our “the spot area” occupancy. Then, our method is validated.

With the above estimation, we can assume that when the V-band flux is the maximum in our observation, we are observing no star spots on the hemisphere of the primary star, and when the V-band flux is the minimal, we are observing the primary star with “the spot area” which covers 25 \% of the hemispherical surface.
The area of 25 \% of the hemisphere surface is calculated to be 20 $R_{\odot}^2$. The values are shown in Table \ref{tab:3}.\footnote{On the contrary, if we assume that the photometric modulation is due to the spots on the secondary star, the spot area is calculated to be 1.5 times larger than the hemisphere of the secondary star.}
Here, we must note that the binary has an area which cannot be seen from us, as shown in Figure 11 “hidden”. However, whether or not this area contains “the spot area” does not affect the discussion that follows.

\section{Discussion} \label{sec:discussion}

\subsection{Relation between X-ray and H$\alpha$ Emission} \label{sec:res:relation}

The X-ray and H$\alpha$ flare emissions of solar and stellar flares are now known to follow common proportional relationships in the $e$-folding times and the radiation energy during a flare (\cite{2022PASJ...74..477K}). 
According to \citet{2022PASJ...74..477K}, the $e$-folding times of the X-ray and H$\mathrm{\alpha}$ light curves in the decaying phase are in agreement for a timescale ranging from 10$^2$ to 10$^5$ secs. On the other hand, for a flare energy range of $10^{29}$--$10^{38}$ erg in the X-ray band, the ratio of the H$\mathrm{\alpha}$-line to bolometric-X-ray emissions is $\sim$0.1. Here, we note that each bolometric-X-ray luminosity was estimated by converting the observed 2--20 keV band emission to that in the 0.1--100 keV band according to the best-fit thin thermal model.

\begin{figure*}[t]
\begin{center}
\includegraphics[width=1.0\linewidth]{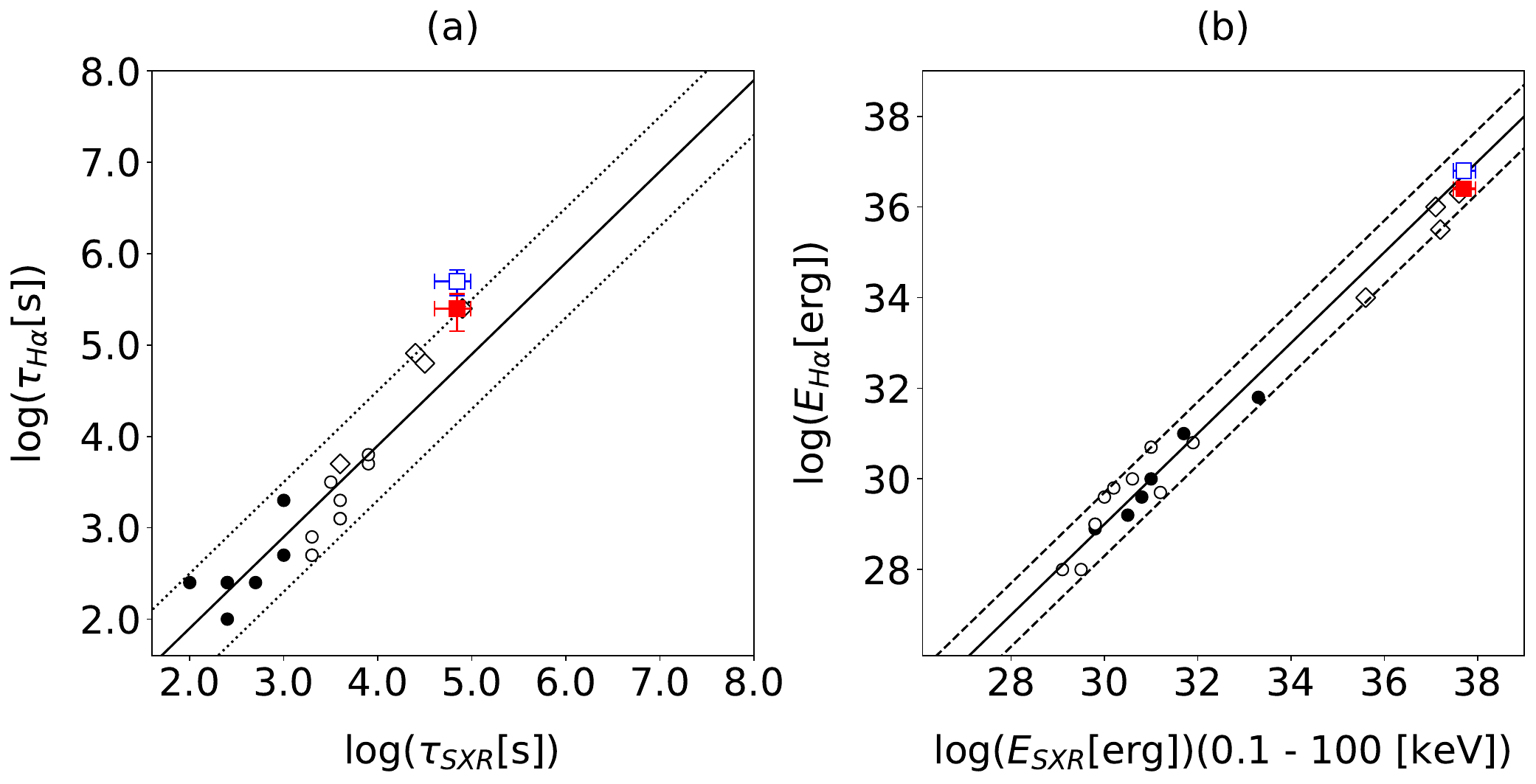}
\end{center}
  \caption{Relations between the properties obtained in the H$\alpha$ line band and that in the X-ray band for solar and stellar flares. The open and filled circles represent data for solar and dMe star's flares, while open diamonds indicate that for RS Cvn stars (see \cite{2022PASJ...74..477K} and the references therein).The filled squares represent our results obtained by the single exponential fitting, and the open squares shows that obtained with the exponential-sinusoidal model fitting (Equation \ref{eq:fit}). The solid line is the best-fit model in \citet{2022PASJ...74..477K}, and the dotted lines indicate 1.6 times of the standard deviation  ($\sim$90 \%) of the data points around the model \citet{2022PASJ...74..477K}. (a): {\it e}-folding time in H$\alpha$ and X-ray bands  (b): radiative energies in H$\alpha$ and bolometric X-ray (0.1--100 keV) bands. \\
  Alt text: Two-panel scatter plot comparing H$\alpha$ and X-ray flare properties for different types of stars.}
  \label{fig:ha-x}
\end{figure*}

In Figure \ref{fig:ha-x}, we plotted our results in the same diagrams as in \citet{2022PASJ...74..477K}. Although there is a hint of deviation in the $e$-folding time in Figure \ref{fig:ha-x} (a) for the data which was fitted with the exponential decay and sinusoidal model, our obtained plots are both roughly located on the previous relationships but at the high-end.
The fact that the physical parameters in the two different bands are located in the common proportional relationships means that the emission regions in the two bands are in close proximity.

\subsection{Emission region of H$\alpha$ line} \label{sec:dis:loc}

\begin{figure}[htbp]
\begin{center}
\includegraphics[width=1.0\linewidth]{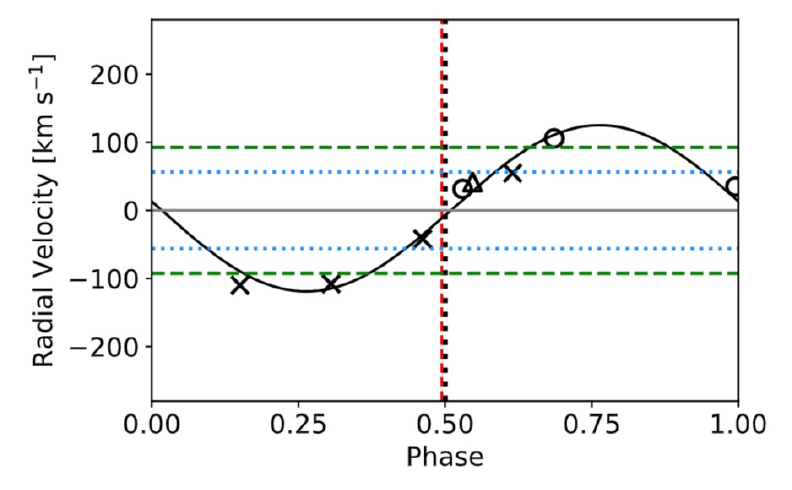}
\end{center}
  \caption{A phase-folded curve of the radial velocity of UX Ari. Orbital period is assumed to be 6.437888 days \citep{2017ApJ...844..115H}. The circle, cross, and triangular marks represent the data for the periods from MJD 59672.4-59675.4, MJD 59675.4-59681.9, and MJD 59681.9-59688.3, respectively. The horizontal dashed and dotted lines correspond to the amplitude for the rotations at ``the edge'' and the center of the primary, respectively. The solid curved line represents the best-fit for a single sinusoidal model. A vertical dashed line indicates the phase when the flare occurred (MAXI trigger time, 2022 April 3 at 04:33:24 UTC), while a vertical dotted line indicated phase of 0.5. \\
  Line graph showing phase-folded variation of radial velocity of UX Ari.}
  \label{fig:rot-RV}
\end{figure}

Figure \ref{fig:rot-RV} presents the RV derived from the central wavelengths (see Section \ref{sec:res:ha}). The data are folded with the spin/orbital period of UX Ari of 6.437888 \citep{2017ApJ...844..115H} as we did in Figure \ref{fig:rot-CAT}. The RV variation can be reproduced with a single sinusoidal curve model with an amplitude of 122 $\pm$ 12 km s$^{-1}$ (the error is 68 \% confidence level). 
The central wavelength of the observed H$\alpha$ line fluctuates centered around 6562.8 {\AA}, the wavelength of the line from the stationary H$\alpha$ source.

Figure \ref{fig:rot-RV} shows that the data points for the secondary bump (crosses) share the same sinusoidal curve as the points in the initial flare phase (circle).
This suggests that the emission at the secondary bump is located in a close area as that in the initial flare.
Although the secondary bump can originate from either (a) a second flare occurring at almost the same location as the first flare, or (b) the same flare reappearing due to stellar rotation, hereafter we assume that it occurred on occasion (b).

\subsubsection{Longitude obtained with phases} \label{sec:dis:loc:longitude}

\begin{figure*}[t]
\begin{center}
\includegraphics[width=1.0\linewidth]{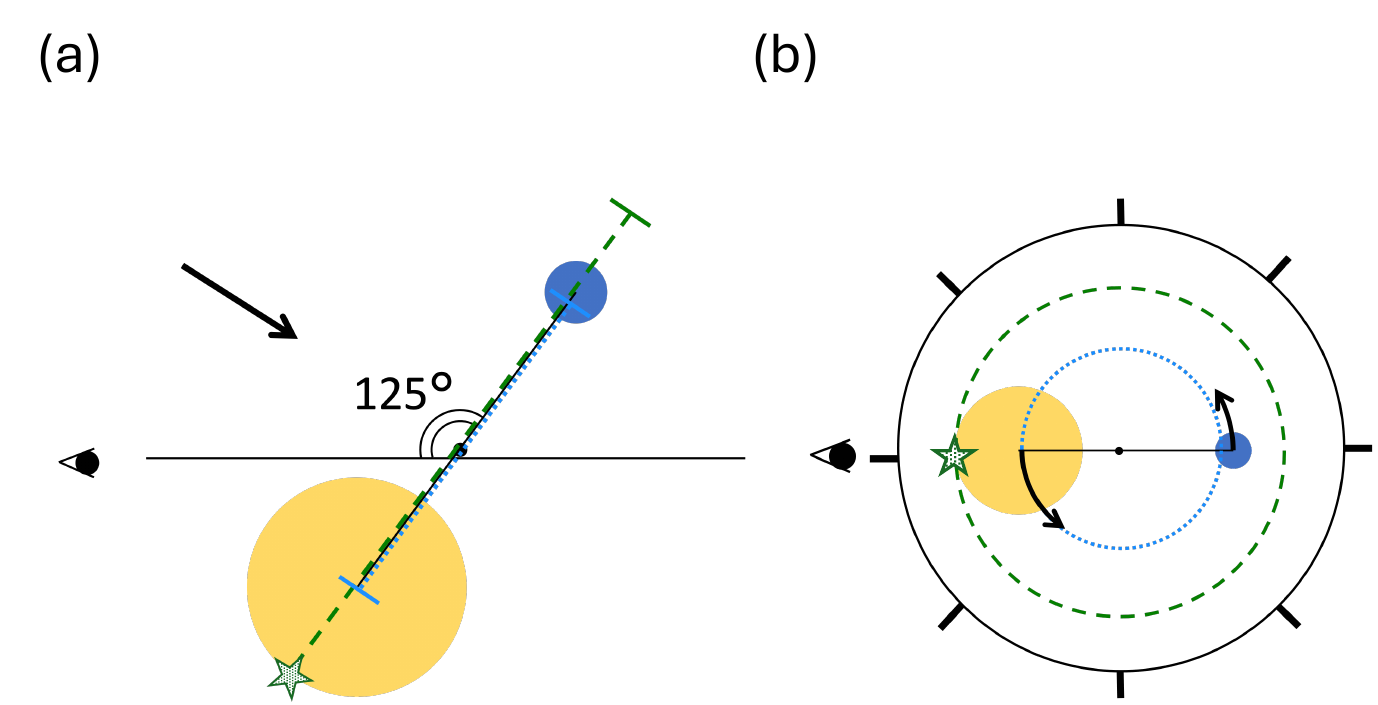}
\end{center}
  \caption{A schematic view showing the configuration between the binary system and the observer (eye marks). The phase 0.5 is assumed. The larger and smaller filled circles represent the primary star (K0IV) and the companion star (G5V), respectively. The orbit of the center of the primary is shown with a dotted line, while that of  ``the edge'' is with a dashed line. (a): A view of a plane which includes the line of the sight, the center of the mass of the binary, the center of the primary, and that of the secondary stars. The arrow indicates the direction vertical to the orbital plane. The star mark on the primary indicates ``the edge'', the most distant point from the center of the mass of the binary.\\ 
  (b): A view from upstream of the arrow depicted in the panel (a).\\
  Alt text: Two-panel schematic diagram showing the geometry of the UX Ari binary system and the observer’s line of sight at orbital phase 0.5.}
  \label{fig:location}
\end{figure*}

With the RV curve obtained (Figure \ref{fig:rot-RV}), we can infer the location of the superflare as follows (see also Figure \ref{fig:location}); 
First, the RV indicates 0 at phase 0.5 and exhibits negative values (blue shift) in phase 0--0.5, while it has positive values (red shift) in phase 0.5--1.0. This indicates that the flare occurred on the side of the primary star and on the plane containing the rotation axis and the center of the primary star, since our definition is the same as that of \citep{2017ApJ...844..115H}. 
This idea is supported by the light curve: we have found that the sinusoidal peaks in Figure \ref{fig:lcfit_ha} (b) exist at phase 0.43 $\pm$ 0.05 (the error is 68 \% confidence level), which is consistent with phase 0.5, the phase at the RV peak. 

Phase 0.5 is also the same timing as “the spot area” appears the largest to us (see Section \ref{sec:res:rotation}). This suggests that the H$\alpha$ line emitting region lies on “the spot area” as its root.

With all of the above information, the longitude of the H$\alpha$ emission region is well-determined. 

\subsubsection{Distance from the rotational axis in rigid rotation} \label{sec:dis:loc:distance}

Figure \ref{fig:location} indicates a schematic picture that shows the configuration between the binary system and the observer. A binary of this type can be considered to be in one rigid rotational motion as a whole, with the center of rotation being the center of mass of the two binary components. Therefore, referring to the RV value and the corresponding phase, it is possible to determine the distance between the emitting region of the H$\alpha$ line and the center of mass of the binary star.

With the above picture described in Figure \ref{fig:location}, the radial velocity should be expressed by Equation \ref{eq:RV}, and then the peak of the radial velocity should be expressed by Equation \ref{eq:RVpeak};

\begin{equation} \label{eq:RV}
R V=\frac{2 \pi D}{P_{rot}} \sin i \sin \left(\frac{2 \pi t}{P_{rot}}+\phi\right)
\end{equation}

\begin{equation} \label{eq:RVpeak}
R V _{peak}=\frac{2 \pi D}{P_{rot}} \sin i 
\end{equation}

Here, $P_{rot}$ is the orbital period, $D$ is the distance from the rotational axis to the H$\alpha$ line emission region, $\phi$ is the initial phase, and $i$ is inclination angle.
Substituting $R V _{peak} = 122$ $\pm$ 12 km s$^{-1}$ and $i = 125$ degree into Equation \ref{eq:RVpeak}, we obtain $D = 19 \pm 2$ $R_{\odot}$ (the error is 68 \% confidence level). 
This value is larger than the radius of the green dashed circle (14.4 $R_{\odot}$) shown in Figure \ref{fig:location} (b), which indicates the orbit of the most distant surface of the primary from the rotational axis. In the following, we call this point on the primary ``the edge''. 

\subsubsection{Latitude obtained from appearing and disappearing of “the spot area” and that of the emission region of H$\alpha$ line} \label{sec:dis:loc:latitude}

According to Equation \ref{eq:fit}, 
the emission of the H$\alpha$ line is 100 \% visible at phase 0.5, but only 40 $\pm$ 10 \% (the error is 68 \% confidence level) is visible at phase 0 from our direction.
Here we calculated equation $(1-B)/(1+B)$, substituting the best-fit value for the amplitude $B$ in Equation \ref{eq:fit}, 0.39, into the equation.

In summary, the emission region of the H$\alpha$ line should exist at the location that satisfies this condition and the condition of $D = 19$ $R_{\odot}$ obtained in Section \ref{sec:dis:loc:distance}. A good candidate region exists at 5 $R_{\odot}$ above ``the edge''. We show the location with an orange square in Figure \ref{fig:location3}.

\subsection{Latitude of “the spot area”} \label{sec:spotarea}

\begin{figure}[htbp]
\begin{center}
\includegraphics[width=1.0\linewidth]{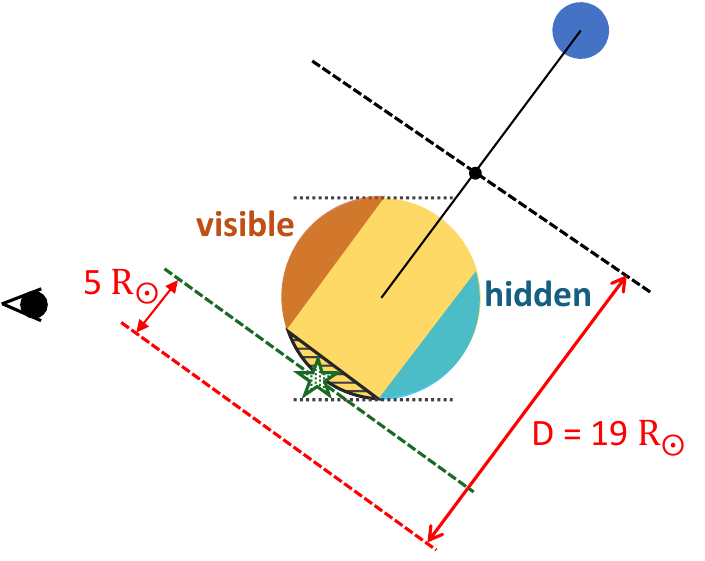}
\end{center}
  \caption{A schematic view showing candidate location for the H$\alpha$ emission region (red broken line) and that for ``the spot area'' (shaded area). This view is written for the phase 0.5. ``The edge'' on the primary is shown with a star mark. Any points on the red broken line have $|R V _{peak}| = 122$ km s$^{-1}$ at phase 0.75 and 0.25, with the distance from the binary's center of mass of 19 $R_{\odot}$. The shaded area is 100 \% visible at phase 0.5 and only 0 \% visible at phase 0, while the “visible” and “hidden” areas shown in this figure are always visible to observers regardless of the orbital phase and areas that are always invisible, respectively.\\
  Alt text: Schematic diagram showing candidate locations of the H$\alpha$ emission region and the spot area on the primary star of UX Ari at orbital phase 0.5.}
  \label{fig:location2}
\end{figure}

As discussed in Section \ref{sec:res:rotation}, “the spot area” is 100 \% visible at phase 0.5 and only 0 \% visible at phase 0. Therefore, the brown and blue regions in Figure \ref{fig:location2}, i.e., “the polar regions”, are excluded as candidates. The remaining latitudes are 100 \% visible at phase 0.5 and completely invisible at phase 0, and then all the remaining latitudes can be regarded as candidates. This region is indicated by the shaded area in Figure \ref{fig:location2}. The latitude covers 70 degree in size, i.e., 20 \% of the circumference of the primary star. If we square the latitude, it covers roughly 20 \% of the hemisphere, and it is roughly the same as the covering factor of “the spot area” (25 \%) we obtained in Section \ref{sec:res:rotation}. Then we can say that the location of “the spot area” is well-determined.

\subsection{Loop Length obtained with X-ray Observations} \label{sec:dis:loop}

As discussed in Section \ref{sec:dis:loc:latitude} and Section \ref{sec:spotarea}, the location of “the spot area” is well-determined to be around ``the edge'' on the primary star, and that of the emission region of H$\alpha$ line is consistent with the location at 5 $R_{\odot}$ above ``the edge''. In contrast, no proof was obtained for the emission of the H$\alpha$ line at the secondary star nor that at the location between the primary star and the secondary star. Independently, as discussed in Section \ref{sec:res:relation}, it is shown that there are two common proportional relationships between two physical parameters in the X-ray and H$\alpha$ emission line bands, suggesting that the emission regions in the two bands are in close proximity. 
With these ideas, the X-ray emission region should also be located near/above ``the edge'', or “the spot area”.

In this subsection, we try to estimate the size of the X-ray emitting region to know whether the root is consistent with “the spot area”.
As shown in Section \ref{sec:res:xray} and Table \ref{tab:2}, we have obtained a best-fit $e$-folding time of $70 \pm 30$ ks by fitting the X-ray light curve in the 2--6 keV band with an exponential decay model. If the cooling is predominantly radiative, then this can be used as an approximation of the radiative cooling timescale $\tau$, which is given by $\tau=3 n_{e} k T / n_{e}^{2} \Lambda(T)$, where $n_{e}$ is the electron density and $\Lambda(T)$ is the emissivity per electron. We define $l_{\rm X}$ as the loop length obtained from this method. At a temperature $T = 5_{-2}^{+3} \times 10^{7}$ K and $\tau = 70 \pm 30$ ksec, the derived electron density is $n_{e}=2 \pm 1 \times 10^{10}$ $\mathrm{cm}^{-3}$. Here and hereafter, all errors for the physical parameters are indicated for the 1 sigma error. This density, along with the emission measure derived from spectral fits, gives an emission volume of $1.2 \pm 0.86 \times 10^{35}$ $\mathrm{~cm}^{3}$. 
If the plasma shape is similar to solar loops (i.e., the aspect ratio, $\alpha =$ loop diameter/a loop length, is 0.1--1; \cite{1989A&A...213..245V}; \cite{1995PASJ...47..251S}) and the cross section is constant, then the loop length and the diameter of the loop are $7 \pm 2 ~R_{\odot}$ and $7\pm 2 ~R_{\odot}$ when $\alpha = 1$, and $40 \pm 10 ~R_{\odot}$ and $4 \pm 1 ~R_{\odot}$ when $\alpha = 0.1$. The estimated properties for the regions emitting X-rays and H$_\alpha$ and the size of “the spot area” are compiled in Table \ref{tab:3}.

\begin{table*}
\caption{Estimated properties for the X-ray and H$\alpha$ emitting regions, and the size of "the spot area". \label{tab:3}}
{\centering
\begin{tabular*}{\textwidth}{@{\extracolsep{\fill}} ccccccccc c}
\hline
$n_{e, X}$ & $S_{X, 1}$ & \multicolumn{2}{c}{$l_{\rm X, 1}$} & $S_{X, 0.1}$ & \multicolumn{2}{c}{$l_{\rm X, 0.1}$} & \multicolumn{2}{c}{$l_{H\alpha}$} & $S_{spot}$ \\
\cmidrule(lr){3-4}
\cmidrule(lr){6-7}
\cmidrule(lr){8-9}
$[10^{10}$ cm$^{-3}]$ & $[R_{\odot}^{2}]$ & $[10^{11}$ cm$]$ & $[R_{\odot}]$ & $[R_{\odot}^{2}]$ & $[10^{11}$ cm$]$ & $[R_{\odot}]$ & $[10^{11}$ cm$]$ & $[R_{\odot}]$ & $[R_{\odot}^{2}]$ \\
\midrule
2 & 50 & 5 & 7 & 10 & 20 & 40 & 3 & 5 & 20 \\
(1--3) & (30--70) & (4--6) & (5--9) & (2--20) & (10--30) & (20--50) & (2--4) & (2--6) & \\
\bottomrule
\end{tabular*}
\par
}
\vspace{2mm}
\begin{flushleft}
\textbf{Note. } $n_{e, X}$ is the electron density calculated assuming that cooling is predominantly radiative. 
$S_{X, 1}$ and $S_{X, 0.1}$ are the constant cross sections when the plasma shape is pole with aspect ratio, loop diameter / a loop length, of 1 or 0.1, respectively.
$l_{\rm X, 1}$ or $l_{\rm X, 0.1}$ is the loop lengths when the aspect ratio of 1 or 0.1, respectively.
$l_{\rm H\alpha}$ is the H$\alpha$ loop length estimated by radial velocity. $S_{\rm spot}$ is "the spot area" diameter estimated by V-band variation in the quiescent phase.
The errors are 68 \% confidence range.
\end{flushleft}
\end{table*}

In this paper, we will not discuss the loop length calculated using the Shibata \& Yokoyama method \citet{1999ApJ...526L..49S,2002ApJ...577..422S}, because the preflare density required by this method is known to have a range of three orders of magnitude from the observations, and the loop length is not easily constrained.
In addition, the preflare density includes a density range that does not satisfy the condition Tc$<$Ta, which is assumed by Shibata \& Yokoyama \citet{1999ApJ...526L..49S,2002ApJ...577..422S}.

\subsection{A picture of the flaring region obtained with the simultaneous and monitoring observations} \label{sec:dis:picture}

\begin{figure*}[t]
\begin{center}
\includegraphics[width=1.0\linewidth]{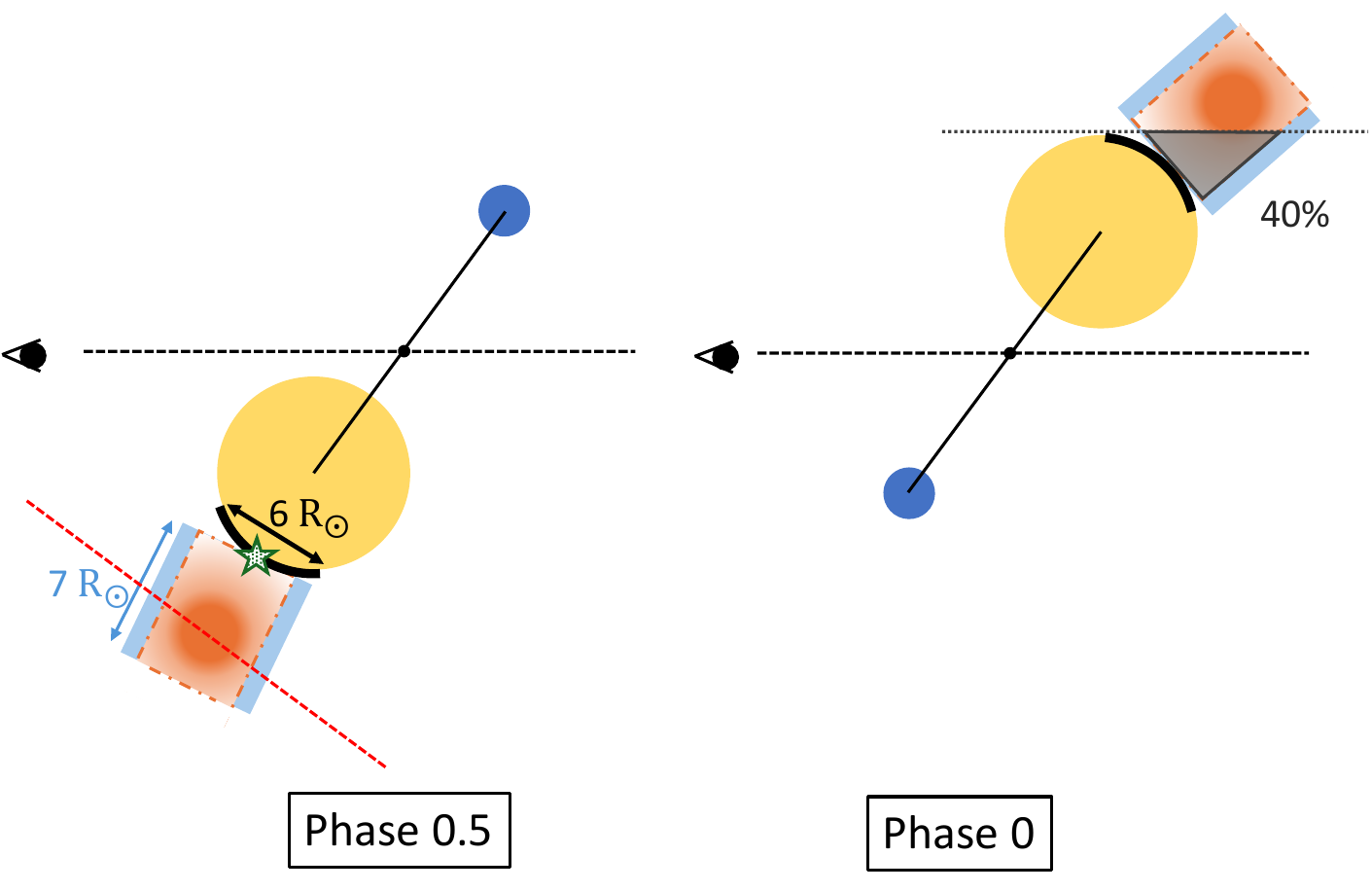}
\end{center}
  \caption{The most plausible picture of the H$\alpha$ line emission region (orange region) considered from the appearing and disappearings of H$\alpha$ line the emission. 
  H$\alpha$ emission region can exist at 5 $R_{\odot}$ above ``the edge'', which is included  in ``the spot area'' shown with the black arc. At phase 0, the orange region is hidden behind the primary star by about 40 \%, as observed. The blue square represents the X-ray emission region, which is assumed to have the aspect ratio of 1.\\
  Alt text: Schematic diagram showing the inferred H$\alpha$ and X-ray emission regions of UX Ari at different orbital phases.}
  \label{fig:location3}
\end{figure*}

With all the above discussions, the following scenario would be the most plausible; (1) the superflare occurred on ``the spot area'' on the primary star which covers about 25 \% of the hemisphere and is located on the opposite side of the secondary star. (2) ``the spot area'' and then the flaring region is located at low latitude. (3) the low latitude of the flaring region causes appearing and disappearing due to stellar rotation, too. (4) The H$\alpha$ emission region is not larger than the stellar size. We show a schematic picture that meets these conditions in Figure \ref{fig:location3}.

We should note that Neidhöfer et al. (1993), Elias et al. (1995), and Torricelli et al. (1995) showed that flares in UX Ari tend to occur more frequently when the primary star is positioned closer to our line of sight. This finding is consistent with the flare site that we identified in this study. In addition, we successfully constrained the latitude of the flare site and the location of the H$\alpha$ emission region.

\section{Summary} \label{sec:summary}

\begin{itemize}
\item[(1)] A giant X-ray flare was detected from an RS CVn-type binary, UX Ari, with MAXI's all sky survey. We performed follow-up optical spectroscopic observations of H$\alpha$ line with the 3.8 m Seimei telescope for 12 days.

\item[(2)] The light curve in the X-ray band shows gradual decay, while that in the H$\alpha$ emission line band shows additional sinusoidal fluctuation on an exponential decay, which is consistent with the recurrence of the flare location due to the binary rotation. The {\it e}-folding decay times in both bands are located at the high-end of the common proportional relationship derived by \citet{2022PASJ...74..477K}.

\item[(3)] From the obtained peak luminosity and the {\it e}-folding time, the radiative energy in the X-ray (the 0.1--100 keV)  and the H$\alpha$ bands are respectively obtained. The energies follow the empirical linear relationship of X-ray and H$\alpha$ flares energy for solar and stellar flares \citep{2022PASJ...74..477K}, giving 
 one of the largest ever recorded among stellar flares observed in both X-ray and H$\alpha$ bands. 

 \item[(4)] The RV variation of the H$\alpha$ emission line has the same period as the orbital period of the binary. The timing and the amplitude of the RV variation indicate the H$\alpha$ emission region is 19 $R_{\odot}$ separated from the axis of the rigid rotation and located 
at the primary star side on the plane containing the rotation axis and the center of the primary star. The region is even farther than the most distant location on the primary star from the rotation axis, “the edge”, which has the distance of 14.4 $R_{\odot}$ from the rotational axis.

\item[(5)] Like the RV, the light curve in the H$\alpha$ emission line also varies with a periodicity consistent with the orbital period, peaking at phase 0.5, when “the edge” is facing the observer.
The minimum value of the H$\alpha$ emission intensity is 40\% of the maximum value, indicating that 60\% of the H$\alpha$ line emission is hidden at phase 0.
These results suggest that the H$\alpha$ emission region is comparable in size to the primary star, and that the H$\alpha$ emission region is located at a relatively low latitude where it can be obscured.

\item[(6)] The optical photometric observations with CAT show periodic rotational modulations, anti-correlated with the light curve in the H$\alpha$ emission line band. This is considered to be the recurrence of “the spot area” due to the binary rotation.
The location is consistent with “the edge”, and the size is estimated to be 25 \% of the stellar hemisphere. This size is reasonable as the root of the H$\alpha$ emission line region.

\item[(7)] If we assume the radiatively cooling plasma, the electron density and the volume of the plasma are derived to be of the order of 10$^{10}$ cm$^{-3}$ and $1 \times 10^{35}$ cm$^3$, respectively, from the X-ray peak spectrum and the light curve. 
When the plasma shape is similar to solar loops with the aspect ratio, $\alpha =$ loop diameter/a loop length, of 0.1--1 with the constant cross section, then the loop length and the diameter of the loop are $7 \pm 2 ~R_{\odot}$ and $7\pm 2 ~R_{\odot}$ when $\alpha = 1$, and $40 \pm 10 ~R_{\odot}$ and $4 \pm 1 ~R_{\odot}$ when $\alpha = 0.1$.

\end{itemize}


\section*{Acknowledgments}
This research has made use of the MAXI data, provided by the RIKEN, JAXA, and MAXI teams. The spectroscopic data used in this paper were obtained through the program 22A-N-CT03 with the 3.8m Seimei telescope, which is located at Okayama Observatory of Kyoto University. We thank 
This work was supported by JSPS KAKENHI Grant Nos. JP21J00316 (K.N.), 25K01041, 24H00248, 24K00680 (K.N., H.M.), 20K04032, 20H05643, 21H01131, 24K00685 (H.M.) and JP21J00106 (Y.N.).
Y.N. was supported from the NASA ADAP award program Number 80NSSC21K0632, and Y. T. was 
supported by Chuo University Grant for Special Research.

\clearpage

\bibliography{UXAri_flare_20220403}{}
\bibliographystyle{aasjournal}

\end{document}